\documentclass[a4paper,11pt]{article}
\usepackage{pos}

\usepackage{algorithm}
\usepackage{algpseudocode}

\newcommand{\bbC}{\mathbb{C}}
\newcommand{\bbR}{\mathbb{R}}

\newcommand{\calD}{\mathcal{D}}
\newcommand{\calF}{\mathcal{F}}
\newcommand{\calJ}{\mathcal{J}}

\newcommand{\calO}{\mathcal{O}}
\newcommand{\calP}{\mathcal{P}}
\newcommand{\calR}{\mathcal{R}}

\newcommand{\ReS}{\mathrm{Re}\,S}
\newcommand{\ImS}{\mathrm{Im}\,S}

\newcommand{\circdot}[1]{\overset{\circ}{#1}}

\newcommand{\vev}[1]{\langle #1 \rangle}

\newcommand{\tr}{\mathrm{tr}\,}
\newcommand{\re}{\mathrm{Re}\,}

\newcommand{\LieG}{\mathfrak{g}}
\newcommand{\LieGC}{\mathfrak{g}^\bbC}

\newtheorem{thm}{Theorem}

\makeatletter
\@addtoreset{equation}{section}

\def\@seccntformat#1{\csname the#1\endcsname.~~}
\makeatother

\title{
Applying the Worldvolume Hybrid Monte Carlo method to lattice gauge theories%
\footnote{Report No.: KUNS-3096}
}
\ShortTitle{WV-HMC for lattice gauge theories}

\author*[a]{Masafumi Fukuma}

\affiliation[a]{Department of Physics, Kyoto University,\\
  Kyoto 606-8502, Japan}

\emailAdd{fukuma@gauge.scphys.kyoto-u.ac.jp}

\abstract{
The numerical sign problem remains one of the central challenges in computational physics. The Worldvolume Hybrid Monte Carlo (WV-HMC) method has recently been proposed as a reliable and computationally efficient algorithm that crucially avoids the ergodicity issues inherent in Lefschetz-thimble approaches. In these proceedings, after outlining the key ideas behind WV-HMC, we present its extension to group manifolds. This provides a rigorous framework for applying WV-HMC to lattice gauge theories.
}

\FullConference{The 42nd International Symposium on Lattice Field Theory (LATTICE2025)\\
2-8 November 2025\\
Tata Institute of Fundamental Research, Mumbai, India\\}

\begin{document}

\maketitle

\section{Introduction}
\label{sec:introduction}

The numerical sign problem is a major obstacle 
to first-principles computations of various physically important systems, 
including finite-density QCD, finite-$\theta$ Yang-Mills theory, 
strongly correlated electron systems, 
and real-time dynamics of quantum many-body systems. 

For the last fifteen years, 
there have been attempts to construct a versatile solution to the sign problem, 
and various methods have been proposed. 
Among them, 
methods based on Lefschetz thimbles have attracted much attention 
because of their mathematical rigor rooted in Picard-Lefschetz theory 
\cite{Witten:2010cx,Cristoforetti:2012su,Cristoforetti:2013wha,
Fujii:2013sra,Fujii:2015bua,Fujii:2015vha,
Alexandru:2015xva,Alexandru:2015sua,Alexandru:2017lqr}. 
There, the integration surface is continuously deformed 
within the complexified space, 
so that the oscillatory behavior is mild 
on the new integration surface. 
It, however, soon became clear that 
the original Lefschetz thimble method generally 
suffers from an ergodicity problem 
\cite{Fujii:2015bua,Fujii:2015vha,Alexandru:2015xva}, 
due to the appearance of zeros of the Boltzmann weight 
on the deformed surface, 
which behave as infinitely high potential barriers 
for the Markov chain walker. 

The first algorithm that simultaneously solves the sign and ergodicity problems 
was the \emph{Tempered Lefschetz thimble (TLT) method} 
\cite{Fukuma:2017fjq,Alexandru:2017oyw}, 
where the deformation parameter $t$ (called the flow time) 
is treated as an extra dynamical variable, 
and the resulting extended configuration space provides a detour 
between two regions that are originally separated by the potential barriers. 
Although the TLT method has proven its versatility and reliability 
in various models \cite{Fukuma:2017fjq,Fukuma:2019wbv}, 
it requires computing the Jacobian of the deformation 
every time two configurations are exchanged between adjacent replicas, 
in order to take into account the difference in volume elements of the replicas. 
The \emph{Worldvolume Hybrid Monte Carlo (WV-HMC) method} 
was then invented to overcome this limitation 
\cite{Fukuma:2020fez} 
(see also Refs.~\cite{Fukuma:2021aoo,Fukuma:2023rrq,Fukuma:2023eru,
Namekawa:2024ert,Fukuma:2025uzg,Fukuma:2025cxg}). 
There, the configuration space is extended to 
a continuous union of deformed surfaces (worldvolume), 
and one considers phase-space integrals 
over the tangent bundle of the worldvolume, 
which carries a natural symplectic structure. 
One no longer needs to compute the Jacobian in configuration generation 
because the phase-space volume element does not change 
in molecular dynamics (MD) 
if one employs a symplectic (and thus volume-preserving) integrator.  
The aim of the present paper is to generalize WV-HMC to group manifolds 
\cite{Fukuma:2025gya}, 
an extension that provides a general setting for lattice gauge theories. 
The presentation closely follows that of Ref.~\cite{Fukuma:2025gya}. 

In the following, 
we write $\vev{X,Y} \equiv \re \tr X^\dagger Y$ 
for matrices $X$ and $Y$.

\section{Complex analysis on complexified groups}
\label{sec:GbbC}

We first define the complexification $G^\bbC$ of a compact Lie group $G$. 
We assume that $G$ is in a faithful unitary representation, 
so that the elements $U_0 \in G$ are expressed by unitary matrices, 
and thus the elements of its Lie algebra $\LieG$ by anti-hermitian matrices. 
We denote a basis of $\LieG$ by $\{T_a\}$ $(a=1,\ldots,N)$, 
which we normalize as 
$\tr T_a T_b = -\delta_{ab}$. 
Accordingly, we raise and lower indices with the rule $A_a = -A^a$.
We introduce the right-invariant Maurer-Cartan 1-form on $G$ by 
\begin{align}
  \theta_0 \equiv dU_0\,U_0^{-1} = T_a \,\theta_0^a
  \quad
  (\text{$\theta_0^a$: real 1-form}),
\end{align}
from which the Haar measure $(dU_0)$ on $G$ is defined as 
\begin{align}
  (dU_0) \equiv \theta_0^1 \wedge \cdots \wedge \theta_0^N.
\end{align}
We let $\LieGC$ be the complexified Lie algebra constructed from $\LieG$, 
and define the complexification $G^\bbC$ of $G$ as%
\footnote{ 
  For $G = SU(n)$ and its Lie algebra $\LieG = \mathfrak{su}(n)$, 
  their complexifications are given 
  by $G^\bbC = SL(n,\bbC)$ and $\LieGC = \mathfrak{sl}(n,\bbC)$. 
} 
\begin{align}
  G^\bbC \equiv \bigl\{
  U = e^{Z} e^{Z'}\cdots e^{Z''} \,|\,
  Z,Z',\ldots,Z'' \in \LieGC \bigr\}.
\end{align}
We introduce the Maurer-Cartan 1-form on $G^\bbC$ by 
\begin{align}
  \theta \equiv dU U^{-1} = T_a \, \theta^a
  \quad
  (\text{$\theta^a$: complex 1-form}),
\end{align}
from which we introduce the holomorphic $N$-form $(dU)$ as 
\begin{align}
  (dU) \equiv \theta^1 \wedge \cdots \wedge \theta^N.
\end{align}
We then have Cauchy's theorem on $G^\bbC$ \cite{Fukuma:2025gya}:
\begin{thm}
  Let $\calD$ be a domain in $G^\bbC$ 
  and $f(U)$ be a holomorphic function on $\calD$. 
  Then, the integral $I_\Sigma$ of $f(U)$ 
  over a real $N$-dimensional oriented submanifold $\Sigma\subset\calD$,
  \begin{align}
    I_\Sigma = \int_\Sigma (dU) \,f(U),
  \end{align}
  depends only on the boundary of $\Sigma$. 
\end{thm}
\noindent
Here, a function $f = f(U)$ is said to be \emph{holomorphic} 
if it depends holomorphically on the matrix elements $U_{ij}$.

\section{WV-HMC for group manifolds}
\label{sec:outline}

\begin{figure}[tb]
  \centering
  \includegraphics[width=50mm]{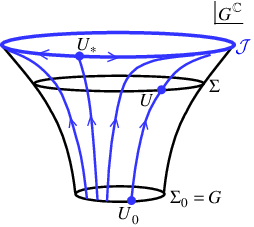}
  \caption{
    Deformation of $\Sigma_0 = G$ 
    into a submanifold $\Sigma$ within $G^\bbC$ \cite{Fukuma:2025gya}. 
    The deformed surface $\Sigma$ approaches a Lefschetz thimble $\calJ$, 
    which consists of points flowing out from a critical point $U_\ast$. 
  }
\label{fig:gt_group}
\end{figure}

Our aim is to numerically evaluate the expectation value 
of an observable $\calO(U_0)$ 
defined by 
\begin{align}
  \vev{\calO} \equiv
  \frac{\int_G (dU_0)\,e^{-S(U_0)}\,\calO(U_0)}
  {\int_G (dU_0)\,e^{-S(U_0)}}. 
\label{vev_G}
\end{align}
We complexify $G = \{U_0\}$ to $G^\bbC = \{U\}$ 
and assume that both $e^{-S(U)}$ and $e^{-S(U)}\,\calO(U)$ 
are holomorphic on $G^\bbC$ 
(which usually holds in cases of physical interest). 
By Cauchy's theorem, 
the expression above can be rewritten 
as a ratio of integrals over a new integration surface $\Sigma$ 
that is obtained by a continuous deformation of $G$ 
(see Fig.~\ref{fig:gt_group}): 
\begin{align}
  \vev{\calO}
  =
  \frac{\int_\Sigma (dU)\,e^{-S(U)}\,\calO(U)}
  {\int_\Sigma (dU)\,e^{-S(U)}}. 
\label{vev_Sigma}
\end{align}
Thus, even when the original path integral on $\Sigma_0=G$ 
suffers from a severe sign problem 
due to the highly oscillatory behavior of $e^{-i\,\ImS(U_0)}$, 
this problem can be significantly alleviated 
if $\ImS(U)$ is almost constant on the new integration surface $\Sigma$. 

In this work, 
we generate such a deformation using the anti-holomorphic gradient flow:
\begin{align}
  \dot{U} = \xi(U)\,U
  ~~\text{with}~~
  U|_{t=0} = U_0.
\label{flow_c}
\end{align}
Here, $\dot{U} \equiv dU/dt$, 
and the drift is given by 
\begin{align}
  \xi(U) \equiv [DS(U)]^\dagger,
\end{align}
where we define the Lie-algebra-valued derivative $DS(U) \in \LieGC$ 
through the variation \cite{Fukuma:2025gya}:
\begin{align}
  \delta S(U) = \tr \bigl[ (\delta U U^{-1})\,DS(U) \bigr]
  = (\delta U U^{-1})^a\,D_a S(U).
\end{align}
This flow equation leads to the monotonicity relation
\begin{align}
  [S(U)]^\centerdot = \tr [(\dot{U}\,U^{-1}) DS(U)]
  =\tr \bigl[ (DS(U))^\dagger\,(DS(U)) \bigr] \geq 0, 
\end{align} 
which shows that 
the real part $\ReS(U)$ always increases along the flow 
[except at critical points where $DS(U)$ vanishes], 
while the imaginary part $\ImS(U)$ remains constant. 
The Lefschetz thimble $\calJ$ associated with a critical point $U_\ast$ 
is defined as the set of points flowing out of $U_\ast$ 
(see Fig.~\ref{fig:gt_group}).  
Since $\ImS(U)$ is invariant along the flow, 
it is constant over $\calJ$. 
Thus, the oscillatory behavior of the integrands in Eq.~\eqref{vev_Sigma}
is expected to be significantly mitigated 
if the integration surface is deformed 
with a sufficiently large flow time $t$ 
so that it reaches the vicinity of $\calJ$.

Sampling on a deformed surface $\Sigma$ 
corresponds to a group-manifold extension of the generalized thimble method 
of Alexandru et al.\ \cite{Alexandru:2015sua}.%
\footnote{ 
  An HMC algorithm on $\Sigma$ in flat space was developed 
  in Refs.~\cite{Alexandru:2019,Fukuma:2019uot} and Ref.~\cite{Fukuma:2023eru}, 
  which we call the \emph{generalized thimble Hybrid Monte Carlo} (GT-HMC). 
  A group-manifold extension of GT-HMC (along with WV-HMC) 
  is presented in Ref.~\cite{Fukuma:2025gya}.
} 
However, as mentioned in Sect.~\ref{sec:introduction}, 
integration over $\Sigma$ will introduce ergodicity issues 
when the flow time is taken to be sufficiently large 
to reduce the oscillatory behavior of $e^{-i\, \ImS(U)}$. 
This motivates us to extend WV-HMC to group manifolds. 

The prescription for introducing WV-HMC to group manifolds 
is the same as in the flat case. 
We first note that 
when we set the deformed surface to $\Sigma=\Sigma_t$ 
(the deformed surface at flow time $t$), 
both the numerator and the denominator of Eq.~\eqref{vev_Sigma} 
do not depend on $t$ due to Cauchy's theorem. 
Thus, we can take averages over $t$ separately 
with an arbitrary common weight $e^{-W(t)}$ 
as in Ref.~\cite{Fukuma:2020fez} 
[we denote $(dU)$ along $\Sigma_t$ by $(dU)_{\Sigma_t}$ 
to specify where it lives]:
\begin{align}
  \vev{\calO}
  =
  \frac{\int_{\Sigma_t} (dU)_{\Sigma_t}\,e^{-S(U)}\,\calO(U)}
  {\int_{\Sigma_t} (dU)_{\Sigma_t}\,e^{-S(U)}}
  =
  \frac{\int dt\,e^{-W(t)}\,\int_{\Sigma_t} (dU)_{\Sigma_t}\,e^{-S(U)}\,\calO(U)}
  {\int dt\,e^{-W(t)}\,\int_{\Sigma_t} (dU)_{\Sigma_t}\,e^{-S(U)}}.
\label{vev_R}
\end{align}
This can be regarded as a ratio of integrals 
over the \emph{worldvolume} $\calR$ defined by 
\begin{align}
  \calR \equiv \bigcup_t \Sigma_t 
  = \{ U(t,U_0) \in G^\bbC \,|\, t \in \bbR,\, U_0 \in G \},
\end{align}
where $U(t,U_0)$ denotes the configuration reached at flow time $t$ 
starting from the initial configuration $U_0$. 
One can effectively constrain the extent of $\calR$ in the $t$-direction 
within a finite interval $[T_0, T_1]$ 
by adjusting $W(t)$ \cite{Fukuma:2023eru}.  
The lower cutoff $T_0$ is chosen  
such that ergodicity issues are absent at $t \sim T_0$, 
while the upper cutoff $T_1$ is chosen 
such that oscillatory integrals are sufficiently tamed at $t \sim T_1$. 
The expectation value \eqref{vev_R} is thus expressed 
as a ratio of the reweighted averages over $\calR$ \cite{Fukuma:2025gya}, 
\begin{align}
  \vev{\calO} 
  &= \frac{\vev{\calF(U)\,\calO(U)}_\calR}
  {\vev{\calF(U)}_\calR}
\\
  \vev{g(U)}_\calR
  &\equiv 
  \frac{ \int_\calR |dU|_\calR\,e^{-V(U)}\,g(U) }
  { \int_\calR |dU|_\calR\,e^{-V(U)} }.
\end{align}
Here, $|dU|_\calR$ is the invariant measure on $\calR$, 
and $V(U)$ and $\calF(U)$ are the potential and the associated reweighting factor 
\cite{Fukuma:2025gya}:%
\footnote{ 
  The function $t(U)$ returns the flow time $t$ 
  for configuration $U=U(t,U_0)$. 
  If we introduce vectors $E_b \equiv T_a E_b^a$ 
  from the Jacobian matrix $E_b^a$ 
  in the linear relation $\theta^a|_{\Sigma_t} = E_b^a\,\theta_0^b$, 
  the reweighting factor can be expressed as 
  $\calF(U) = \alpha^{-1}\, (\det E / \sqrt{\gamma})\, e^{-i\,\ImS(U)}$ 
  with $\gamma_{ab} = \vev{E_a, E_b}$. 
  $\alpha \equiv \sqrt{\vev{\xi_n, \xi_n}}$ 
  is the norm of the normal component 
  $\xi_n \in N_U \Sigma_t$ of $\xi$. 
  The potential $V(U)$ is a real-valued function, 
  and its derivatives are defined by 
  $\delta V = \tr \bigl[ (\delta U U^{-1})\,DV
  + (\delta U U^{-1})^\dagger\,(DV)^\dagger \bigr]$. 
  See Ref.~\cite{Fukuma:2025gya} for details. 
} 
\begin{align}
  V(U) 
  \equiv \ReS(U) + W(t(U)),
  \quad
  \calF(U) 
  \equiv
  \frac{dt\,(dU)_{\Sigma_t}}{|dU|_\calR}\,e^{-i\,\ImS(U)}.
\end{align}

The reweighted averages $\vev{\cdots}_\calR$ 
can be rewritten as integrals over the tangent bundle of $\calR$, 
\begin{align}
  T \calR = \{ (U,\pi)\,|\,
  U \in \calR,\,\pi \in T_U \calR \},
\end{align}
as in the flat case \cite{Fukuma:2023eru} 
(see also Ref.~\cite{Fukuma:2020fez}), 
\begin{align}
  \vev{ g(U) }_\calR 
  &= \frac{\int_{T \calR}\, d\Omega_\calR \,e^{-H(U,\pi)}\,g(U)}
  {\int_{T \calR}\, d\Omega_\calR \,e^{-H(U,\pi)}},
\\
  H(U,\pi)
  &= \frac{1}{2}\,\vev{\pi,\pi} + V(U).
\end{align}
Here, we have introduced a symplectic structure on $T\calR$ 
with the symplectic 2-form 
$\omega_\calR =  d\, \vev{\pi, \theta_\calR}$ 
($\theta_\calR$ denoting $\theta$ along $\calR$), 
and set the symplectic volume form $d\Omega_\calR$ as  
\cite{Fukuma:2025gya}
\begin{align}
  d\Omega_\calR
  = \frac{\omega_\calR^{N+1}}{(N+1)!}.
\end{align}

In the rest of this section, 
we construct a Markov chain on $T\calR$
that has the equilibrium distribution 
$\propto e^{-H(U,\pi)}$. 
To this end, 
we first \emph{define} Hamiltonian dynamics 
on the tangent bundle of $G^\bbC$, $T G^\bbC \equiv 
\{(U,\pi)\,|\,U \in G^\bbC,\,\pi \in T_U G^\bbC \}$, 
using the first-order action 
\cite{Fukuma:2025gya} 
\begin{align}
  I[U(s),\pi(s)] 
  = \int ds\,\bigl[ \vev{\pi,\circdot{U} U^{-1}} - H(U,\pi) \bigr],
\label{action}
\end{align}
where $\circdot{U} \equiv dU/ds$. 
The first term $\vev{\pi,\circdot{U} U^{-1}}$ 
corresponds to the symplectic potential 
$a = \vev{\pi,\theta}$ of the symplectic 2-form 
$\omega = da = d \vev{\pi, \theta}$,  
for which the Poisson brackets take the form \cite{Fukuma:2025gya} 
\begin{align}
  \{U_{ij},\pi^\dagger_{kl}\}
  = 2\,\Bigl( \delta_{il}\,U_{kj} - \frac{1}{n}\,U_{ij}\,\delta_{kl} \Bigr),
  \quad 
  \{\pi_{ij},\pi_{kl}\}
  = 2\,(-\delta_{il}\,\pi_{kj} + \pi_{il}\,\delta_{kj}),
  \quad \cdots
\end{align}
One can check that 
the obtained Hamilton's equations \cite{Fukuma:2025gya}
\begin{align}
  \circdot{U} = \pi \,U,
  \quad
  \circdot{\pi} = -2\, [DV(U)]^\dagger + [\pi,\pi^\dagger]
\label{hamilton_eqs}
\end{align}
are indeed written as 
$\circdot{U} = \{U,H\}$, $\circdot{\pi} = \{\pi,H\}$. 

We then define 
the MD evolution operator of step size $\Delta s = \epsilon$ 
on $T G^\bbC$ as 
\begin{align}
  T \equiv e^{-(\epsilon/2)\{\ast,K\}}\,e^{-\epsilon\{\ast,V\}}\,
  e^{-(\epsilon/2)\{\ast,K\}},
\end{align}
which differs from the continuous evolution operator $e^{-\epsilon\{\ast,H\}}$ 
by $O(\epsilon^3)$. 
A straightforward calculation \cite{Fukuma:2025gya} shows that 
a single MD step 
$(U,\pi) \to (U',\pi') \equiv (T(U),T(\pi))$
is given by
\begin{align}
  \pi_{1/2} &= \pi - \epsilon\,[DV(U)]^\dagger,
\label{md1}
\\
  U' 
  &= e^{\epsilon (\pi_{1/2} - \pi_{1/2}^\dagger)}\,
  e^{\epsilon \pi_{1/2}^\dagger}\,U,
\label{md2}
\\
  \pi'
  &= e^{\epsilon (\pi_{1/2} - \pi_{1/2}^\dagger)}\,\pi_{1/2}\,
  e^{-\epsilon (\pi_{1/2} - \pi_{1/2}^\dagger)}
  - \epsilon\,[DV(U')]^\dagger.
\label{md3}
\end{align}
One can prove \cite{Fukuma:2025gya} that this is 
(a) exactly reversible with 
\begin{align}
  U 
  \leftrightarrow U',
  \quad
  \pi \leftrightarrow -\pi',
  \quad
  \pi_{1/2} 
  \leftrightarrow 
  -e^{\epsilon (\pi_{1/2} - \pi_{1/2}^\dagger)}\,\pi_{1/2}\,
    e^{-\epsilon (\pi_{1/2} - \pi_{1/2}^\dagger)},
\label{reversibility}
\end{align}
(b) symplectic, $\omega' = \omega$, 
and (c) approximately preserving $H(U,\pi)$ 
as $H(U',\pi') = H(U,\pi) + O(\epsilon^3)$. 

Once consistent MD [Eqs.~\eqref{md1}--\eqref{md3}] is defined on $T G^\bbC$, 
constrained MD on $T\calR$ can be constructed 
using the RATTLE algorithm \cite{Andersen:1983,Leimkuhler:1994} 
(see Ref.~\cite{Fukuma:2025gya} for details),%
\footnote{ 
  The gradient of the potential 
  can be set to the following form \cite{Fukuma:2025gya}: 
  $
    [DV(U)]^\dagger = 
    (1/2)\,\bigl[ \xi + (W'(t)/\alpha^2)\,\xi_n \bigr].
  $
} 
\begin{align}
  \pi_{1/2} &= \pi - \epsilon\,[DV(U)]^\dagger - \lambda,
\label{rattle1}
\\
  U' 
  &= e^{\epsilon\,(\pi_{1/2} - \pi_{1/2}^\dagger)}\,
  e^{\epsilon\,\pi_{1/2}^\dagger}\,U,
\label{rattle2}
\\
  \pi'
  &= e^{\epsilon\,(\pi_{1/2} - \pi_{1/2}^\dagger)}\,\pi_{1/2}\,
  e^{-\epsilon\,(\pi_{1/2} - \pi_{1/2}^\dagger)}
  - \epsilon\,[DV(U')]^\dagger - \lambda',
\label{rattle3}
\end{align}
where the Lagrange multipliers $\lambda \in N_U\calR$ 
and $\lambda' \in N_{U'}\calR$ are determined 
such that $U' \in \calR$ and $\pi' \in T_{U'}\calR$, respectively. 
This MD step is again 
(a) reversible 
[Eq.~\eqref{reversibility} 
together with interchange $\lambda \leftrightarrow \lambda'$], 
(b) symplectic, $\omega'_\calR = \omega_\calR$ 
(thus volume preserving, $d\Omega'_\calR = d\Omega_\calR$), 
and (c) approximately preserving $H(U,\pi)$ to the same precision. 

We can now define the Markov chain on $T\calR$ 
as consisting of two stochastic processes \cite{Fukuma:2025gya}:

\noindent
\underline{(1) Heat bath for $\pi$}:
\begin{align}
  P_{(1)}(U',\pi'\,|\,U,\pi) = e^{-(1/2)\,\vev{\pi', \pi'}}\,
  \delta_\calR(U',\, U),
\end{align}
where $\delta_\calR(U',\,U)$ is the delta function on $\calR$.%
\footnote{ 
  When $U \in \calR$ is parametrized as $U = U(t,U_0)$, 
  the delta function is proportional to  
  $\delta(t'-t)\,\delta(U'_0,\, U_0)$, 
  where $\delta(U'_0,\, U_0)$ is the bi-invariant delta function 
  associated with the Haar measure $(dU_0)$ on $G$. 
  Jacobian factors can be neglected 
  in the argument for detailed balance of MD \cite{Fukuma:2025gya}.
} 
$\pi' \in T_U \calR$ can be generated by drawing $\tilde\pi \in T_U G^\bbC$ 
from the Gaussian distribution 
$\propto e^{-(1/2)\,\vev{\tilde\pi,\tilde\pi}}$ 
and projecting it onto $T_U \calR$. 

\noindent
\underline{(2) MD followed by Metropolis test}:%
\footnote{ 
  The transition probability for $(U',\pi') = (U,\pi)$ 
  is determined by the normalization  
  $\int d\Omega'_\calR\, P_{(2)}(U',\pi'\,|\,U,\pi) = 1$. 
} 
\begin{align}
  &P_{(2)}(U',\pi'\,|\,U,\pi) 
\nonumber
\\
  &= 
  \min \bigl(1, e^{-[H(U',\pi') - H(U,\pi)]} \bigr)\,
  \delta_{T\calR}\bigl(
    (U',\pi') ,\, T^{N_\textrm{MD}} (U,\pi)
  \bigr)
  ~~\text{for}~~
  (U',\pi') \neq (U,\pi).
\end{align}
Here, $\delta_{T\calR}(U,\pi)$ is the symplectic delta function 
with respect to the symplectic volume form $d\Omega_\calR$, 
and $N_\textrm{MD}$ is the number of MD steps. 

Since our observables depend only on $U$, 
the above algorithm can be viewed as a stochastic process 
on $U$ alone as in the standard HMC algorithm \cite{Duane:1987de}: 
\begin{itemize}
  \item
  \underline{Step 1 (momentum refresh)}:
  Given $U \in \calR$, 
  generate $\tilde\pi \in T_U G^\bbC$ 
  from the Gaussian distribution 
  $\propto e^{-(1/2)\, \vev{\tilde\pi, \tilde\pi}}$,
  and project it onto $T_U\calR$ 
  to obtain $\pi = \Pi_\calR\, \tilde\pi$.
  
  \item
  \underline{Step 2 (MD)}:
  Evolve $(U,\pi) \to (U',\pi')$ 
  by repeatedly performing the update \eqref{rattle1}--\eqref{rattle3}.
  
  \item
  \underline{Step 3 (Metropolis test)}:
  Accept the proposed $U'$ 
  with probability $\min \bigl(1, e^{-[H(U',\pi') - H(U,\pi)]} \bigr)$.
  
\end{itemize}

\section{Numerical tests: one-site model}
\label{sec:1site}

The one-site model for a compact group $G=SU(n)$ 
is defined by the action 
\begin{align}
  S(U) 
  \equiv \beta e(U) 
  ~~\text{with}~~
  e(U) \equiv -\frac{1}{2n}\,\tr (U + U^{-1}).
\end{align}
We take $\beta$ to be purely imaginary, 
which makes the Boltzmann weight a pure phase factor of constant modulus. 
In the following numerical tests, 
we take $e(U)$ as the observable. 

The variation of the action is given by 
($\calP$ denotes the traceless projector) 
\begin{align}
  \delta S(U) 
  &= -\frac{\beta}{2n}\,\tr [ \delta U U^{-1} (U - U^{-1}) ]
  = - \frac{\beta}{2n}\,\tr [ \delta U U^{-1} \calP (U - U^{-1}) ].
\end{align}
Comparing this with $\delta S(U) = \tr [\delta U U^{-1} DS(U)]$, 
we obtain 
\begin{align}
  DS(U) = -\frac{\beta}{2n}\,\calP (U - U^{-1}),
\end{align}
and therefore 
\begin{align}
  \xi(U) = [DS(U)]^\dagger 
  = -\Bigl[ \frac{\beta}{2n}\,\calP (U - U^{-1}) \Bigr]^\dagger.
\end{align}
This defines the flow of a configuration,
\begin{align}
   \dot{U} = \xi(U)\,U
   ~~\text{with}~~
   U|_{t=0} = U_0.
\label{flow_c_1site}
\end{align}
We numerically integrate the flow equation 
using an adaptive version 
of the Runge-Kutta-Munthe-Kaas algorithm \cite{RKMK1,RKMK2}.

\begin{figure}[tb]
  \centering
  \includegraphics[width=65mm]
  {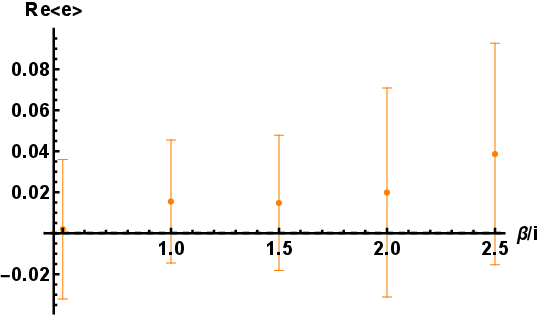}
  \hspace{10mm}
  \includegraphics[width=65mm]
  {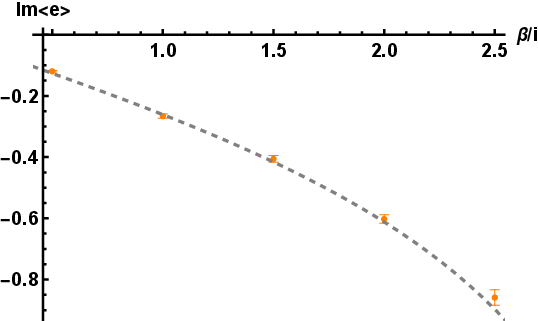}
\\
\vspace{5mm}
  \includegraphics[width=65mm]
  {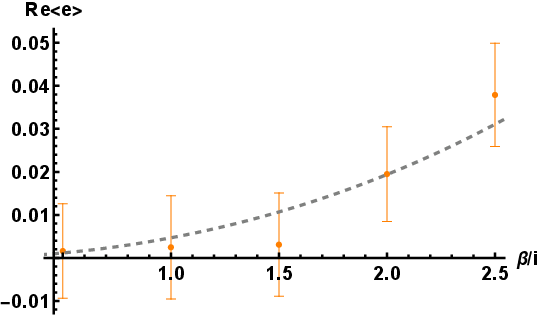}
  \hspace{10mm}
  \includegraphics[width=65mm]
  {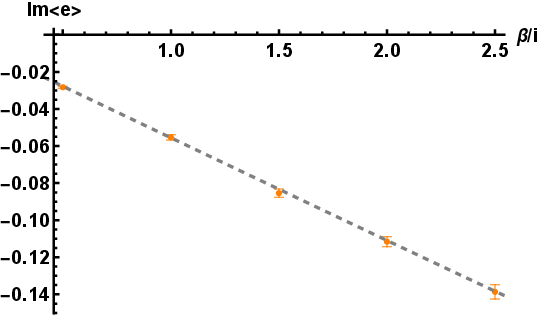}
  \caption{
    Real and imaginary parts of $\vev{e}$ in the one-site model 
    for $\beta \in i\,\bbR$  
    with $G = SU(2)$ [top] and $G = SU(3)$ [bottom] \cite{Fukuma:2025gya}.
    The dashed lines represent the analytical results 
    (for $G = SU(2)$, 
    $\vev{e} = -I_2(\beta)/I_1(\beta)$, 
    for which $\re \vev{e} = 0$).
  }
  \label{fig:energy_density}
\end{figure}

To estimate the observable, 
we introduce boundaries at $T_0 = 0$ and $T_1 = 0.5$ 
and set the MD step size 
to $\Delta s = \epsilon = 0.01$ 
with $N_\textrm{MD} = 50$ steps per trajectory. 
We generate 5500 configurations using WV-HMC, 
with the first 500 configurations discarded.
Figure~\ref{fig:energy_density} shows 
the real and imaginary parts of the energy density $\vev{e}$ 
for various values of $\beta \in i\,\bbR$ \cite{Fukuma:2025gya} 
with $G=SU(2)$ [top] and $G=SU(3)$ [bottom]. 
The results are in good agreement with the analytical values. 
%

\section{Conclusions and outlook}
\label{sec:conclusion}

We have demonstrated that the WV-HMC algorithm 
can be extended to group manifolds \cite{Fukuma:2025gya} 
in such a way that reversibility, symplecticity, 
and approximate energy conservation are all realized 
as in the standard HMC algorithm using the leapfrog integrator. 
The key ingredient is to formulate the algorithm 
using phase-space integrals over the tangent bundle of the worldvolume, 
which naturally carries a symplectic structure.
We have validated the correctness of the algorithms 
through numerical simulations of the one-site model. 

The present formalism can be directly applied to lattice gauge theories 
without any modification to the algorithmic structure. 
The compact group $G$ becomes a product group 
$G = \prod_{x,\mu} G_{x,\mu}$ 
[e.g., $G_{x,\mu} = SU(n)$ at each link $(x,\mu)$], 
and the corresponding Lie algebra is given by 
$\LieG = \bigoplus_{x,\mu} \LieG_{x,\mu} 
= \bigoplus_{x,\mu,a} \bbR\,(T_{x,\mu})_a$ 
with the commutation relations 
$[(T_{x,\mu})_a, (T_{y,\nu})_b] = \delta_{x y}\,\delta_{\mu \nu}\,
C_{ab}{}^c\,(T_{x,\mu})_c$. 
A study of lattice gauge theories with complex actions is now in progress 
and will be reported in forthcoming publications.

\section*{Acknowledgments}
The author thanks Sinya Aoki, Ken-Ichi Ishikawa, Issaku Kanamori, 
Anthony D.\ Kennedy, Yoshio Kikukawa and Yusuke Namekawa 
for valuable discussions. 
This work was partially supported by JSPS KAKENHI 
(Grant Numbers JP20H01900, JP23H00112, JP23H04506, JP25H01533); 
by MEXT as ``Program for Promoting Researches on the Supercomputer Fugaku'' 
(Simulation for basic science: approaching the new quantum era, JPMXP1020230411); 
and by SPIRIT2 2025 of Kyoto University. 


\appendix



\end{document}